\documentclass[conference]{IEEEtran}
\IEEEoverridecommandlockouts
\AtBeginDocument{%
  }

\usepackage{float}
\usepackage{graphicx}
\begin{document}

\title{A Web-Based IDE for DevOps Learning in Software Engineering Higher Education}

\author{Ganesh Neelakanta Iyer, \IEEEmembership{Senior Member, IEEE}, Andrew Goh Yisheng, Metilda Chee Heng Er, \\ Weng Xian Choong and Shao Wei Koh {\thanks{Ganesh Neelakanta Iyer is with Department of Computer Science, National University of Singapore (e-mail: gni@nus.edu.sg), Andrew Goh Yisheng, Metilda Chee Heng Er, Weng Xian Choong and Shao Wei Koh are with School of Continuing and Lifelong Education, National University of Singapore (e-mail: e0493359@u.nus.edu, metildachee@gmail.com, wengxianchoong@gmail.com, kshaowei95@gmail.com).
}}}


\maketitle
\begin{abstract}
DevOps can be best explained as people working together to conceive, build and deliver secure software at top speed. DevOps practices enable software development (dev) and operations (ops) teams to accelerate delivery through automation, collaboration, fast feedback, and iterative improvement. It is now an integral part of the information technology industry, and students should be aware of it before they start their careers. However, teaching DevOps in a university curriculum has many challenges as it involves many tools and technologies. This paper presents an innovative online Integrated Development Environment (IDE) designed to facilitate DevOps learning within university curricula. The devised tool offers a standardized, accessible learning environment, equipped with devcontainers and engaging tutorials to simplify learning DevOps. Research findings highlight a marked preference among students for self-paced learning approaches, with experienced DevOps practitioners also noting the value of the tool. With barriers such as limited hardware/software access becoming evident, the necessity for cloud-based learning solutions is further underscored. User feedback emphasizes the tool's user-friendliness and the imperative of automated installation procedures. We recommend additional exploration into the tool's extensibility and potential for continuous improvement, especially regarding the development of Dev Containers. The study concludes by emphasizing the pivotal role of practical learning tools in the dynamic field of DevOps education and research.
\end{abstract}




%

\section{Introduction}

\begin{figure*}[h]
\centering
\includegraphics[width=\textwidth]{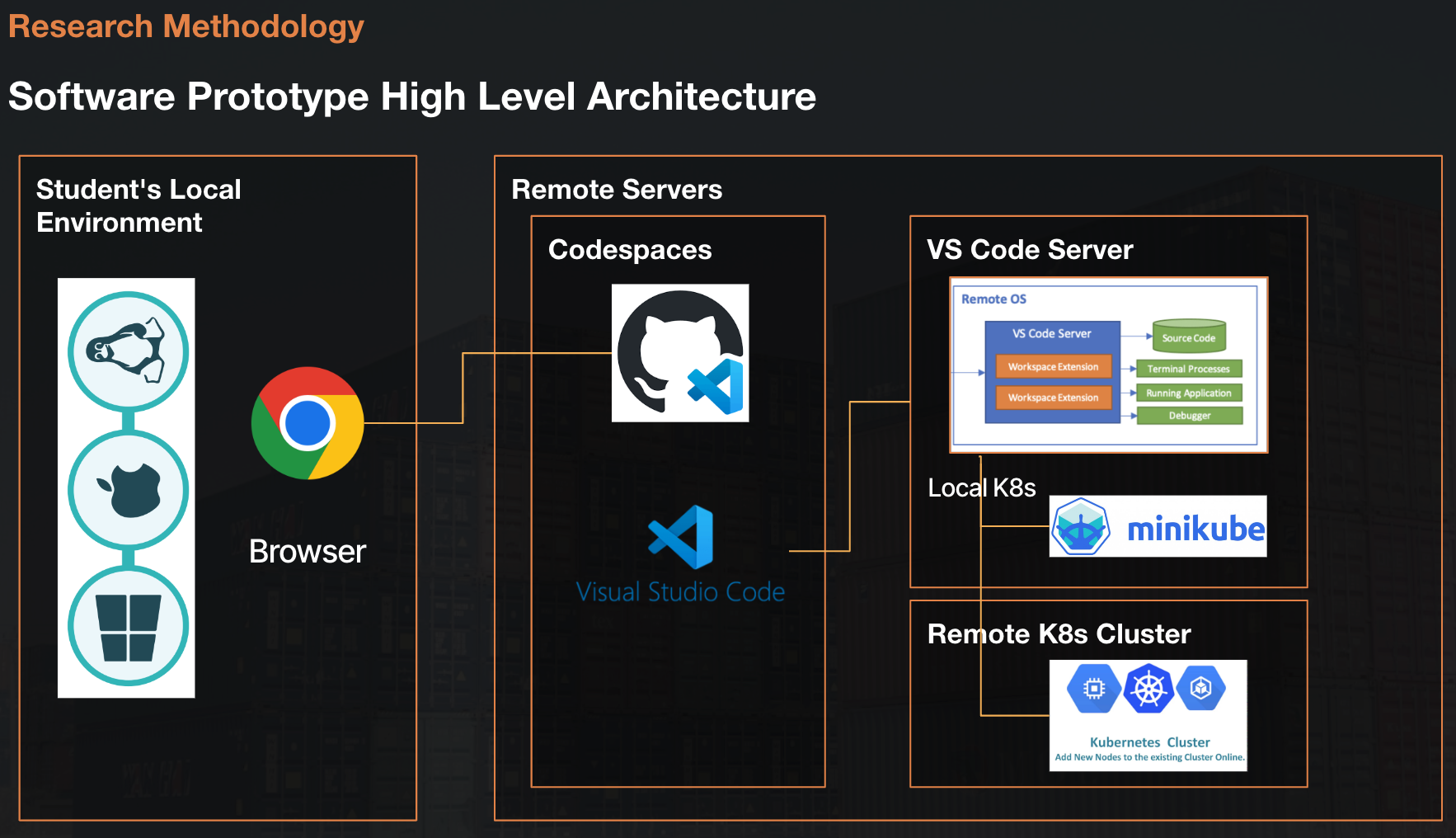}
\caption{Architecture Diagram of ContiNUSd Application}
\label{Architecture Diagram}
\end{figure*}
DevOps, a methodology bridging Development and Operations, is essential for modern software development practices, promoting collaboration, continuous integration, and automated deployment \cite{SystematicMapping}. Its growing importance in the industry necessitates effective educational strategies to equip aspiring professionals with practical skills, particularly in using tools like Kubernetes and Docker, which have been shown to enhance learning and provide hands-on experience \cite{TwoUniversities}.

At Imperial College London, a specialized DevOps Laboratory exercise (DevOps Lab) is integral to preparing students for careers in Software Engineering. Early introduction of DevOps practices within degree programs not only aids in mastering the software system lifecycle but also improves students' adaptability to new tools and documentation, facilitating a smoother transition to professional settings \cite{ICLDevOpsLab}.

Despite its growing prominence, DevOps often receives insufficient coverage in traditional software engineering curricula, which primarily emphasize the initial and intermediate phases of the software lifecycle while neglecting the crucial production phase. This imbalance results in a curriculum that is heavily "Dev"-oriented, lacking adequate focus on the "Ops" component \cite{ACMTeachingDevOpsStoryTelling}.

Addressing this gap, educators are increasingly seeking to integrate DevOps into their programs, albeit amidst challenges. As outlined by Christensen \cite{ACMTeachingDevOpsStoryTelling}, these challenges encompass educators' experience, skills amalgamation, skills emphasis, realistic environment simulation, and evaluation strategies. Christensen further categorizes these challenges into skills acquisition, requiring a shift from knowledge accumulation to skill development, and technical environment, highlighting the need for robust platforms and effective evaluation strategies to assess students accurately and efficiently \cite{ACMTeachingDevOpsStoryTelling}.

\section{Research Overview}

Our research focused on the development of a streamlined and user-friendly learning tool for DevOps education. This tool, designed with the dynamic and evolving landscape of DevOps in mind, aimed to simplify the installation process for both students and educators while ensuring adaptability to industry advancements. With a foundation built on extensibility and an open-source framework, it was conceived to accommodate continuous improvements and the integration of emerging technologies. It was not merely a learning aid but a progressive platform designed to evolve alongside the DevOps industry. The goal was to provide an engaging and enriched learning experience, with a focus on ease of use and preparedness for future technological developments.

\subsection{Research Questions}

Our study explores the effective incorporation of DevOps into IT-related university courses, pivoting around two research questions:

\begin{enumerate}
\item \textbf{Impact of Simplified Tool Setup:} Does the simplification of the setup process for DevOps tools enhance the teaching and learning experience, making it more effective and responsive to the rapid changes in the industry?
\item \textbf{Efficacy of Hands-On Learning Tools:} How effective is a self-guided, experiential learning tool in imparting essential skills and concepts related to containerization management?
\end{enumerate}

To explore these, we crafted a standardized learning environment with an Integrated Development Environment (IDE) and Kubernetes runtime, aiming to evaluate the effects of simplified tool setup and the efficacy of hands-on learning in teaching containerization management.

\section{Research Methodology}

The methodology adopted for our research was a multi-step, iterative process designed to test and refine our DevOps learning tool.

\subsection{Literature Review}

This section reviews the literature on educational tools and methodologies in DevOps to understand the current landscape and identify gaps that our tool aims to fill.

\subsubsection{Case Study: CMU and TUB}
\paragraph{Insights}
A seminal contribution to the literature is the 2021 paper, 'Teaching DevOps: A Tale of Two Universities' by Carnegie Mellon University (CMU) and Technische Universität Berlin (TUB), outlining their experiences in delivering DevOps education. Notably, it advocates for the inclusion of Containers and Containers Management content in the curriculum, covering five out of the eight critical areas in the DevOps Process, with an emphasis on Kubernetes and Docker for enhanced learning and skill acquisition \cite{TwoUniversities}.

\paragraph{Value of DevOps Education: Student Perspectives}
Students from CMU and TUB acknowledged the significant value of DevOps courses in their professional development. The courses provided relevant skills and facilitated smooth transitions into professional roles, as reported by students from both institutions.

\paragraph{Challenges in DevOps Education}
Rapid changes and the emergence of new tools in the DevOps field present challenges for educators in maintaining current and relevant curricula. The dynamic nature of the field requires a learning tool that can adapt quickly to these changes.

\subsubsection{Case Study: Imperial College London}
\paragraph{Insights}
Chatley and Procaccini discussed their approach to DevOps education at Imperial College London, emphasizing the training of students to efficiently use and understand documentation and settings of selected tools \cite{ICLDevOpsLab}.

\paragraph{Positive Learning Outcomes}
Despite initial challenges, students showed improvement in navigating documentation and using new tools, attributing their success to prior experience with similar services. Participants also reported that the skills acquired during the course facilitated their onboarding process at their respective companies and integration into professional teams, highlighting the practical applicability of the course.

\subsubsection{Teaching DevOps Challenges}
\paragraph{Addressing Identified Challenges in DevOps Education}
In the paper "Overcoming Challenges in DevOps Education
through Teaching Methods", the different teaching methods were analysed. However, emphasis was placed on "more active and practice-orientated approaches".

Christensen identified five key challenges in teaching DevOps in the paper titled "Teaching DevOps and Cloud Computing Using a Cognitive Apprenticeship and Story-Telling Approach" \cite{ACMTeachingDevOpsStoryTelling}. Similarly, numerous teams shared similar obstacles \cite{Fernandes_Ferino_Fernandes_Kulesza} \cite{Fernandes}. Our proposed tool addresses three significant challenges pinpointed in this foundational work: 

\begin{itemize}
    \item \textbf{Supporting Teacher Experience:} The tool assists educators, who often have varied levels of expertise in Development (Dev) and Operations (Ops), in delivering and maintaining an up-to-date and relevant curriculum despite the fast-paced evolution of the DevOps field.
    \item \textbf{Emphasizing Skill Acquisition:} Designed to facilitate hands-on experience, the tool allows students to engage actively in creating deployments on Kubernetes, providing theoretical knowledge and practical skills.
    \item \textbf{Providing a Realistic Environment:} The tool incorporates open-source, industry-standard software engineering and DevOps tools, offering students a practical and scalable learning environment that can be adjusted for different levels of system complexity.
\end{itemize}

\subsection{Survey Research}

We wanted to find out if the convenient setup of devops tools able to make teaching DevOps more effective.  
\begin{itemize}
    \item Does this help educators to keep teaching materials up-to-date with industry's fast moving pace?
\item By lowering the barrier to start, will students and educators both be more interested to starting learning DevOps in higher education?

\end{itemize}
Our hypothesis is \textit{Educators can keep pace with latest technology easily by having the automated setup of standard Kubernetes environments be maintained by the open-source community.
}\\ \\
Our second research question was, "Does a self-guided, hands-on learning tool help students to learn DevOps effectively?" We specifically wanted to find out,
\begin{itemize}
\item Can we teach containerisation management concepts and skills effectively? 
\item Are students interested in learning with this tool?
\item Do students prefer self-guided or collaborative learning, especially since DevOps encourages collaboration?
\end{itemize}
Our hypothesis is, \textit{An open-source, self-guided learning tool, with the automated installation of tools setup will encourage students to learn DevOps and help them learn containerisation management skills and concepts effectively.
} \\ \\
\begin{figure}
\centering
\includegraphics[width=3in]{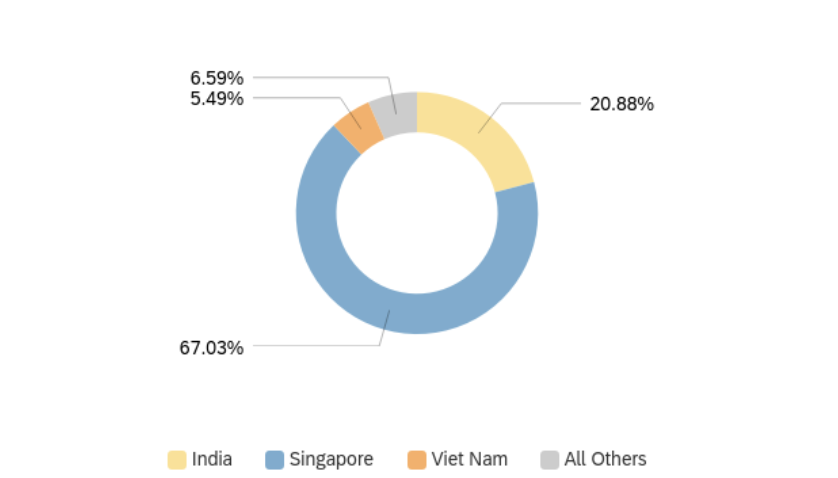}
\includegraphics[width=3in]{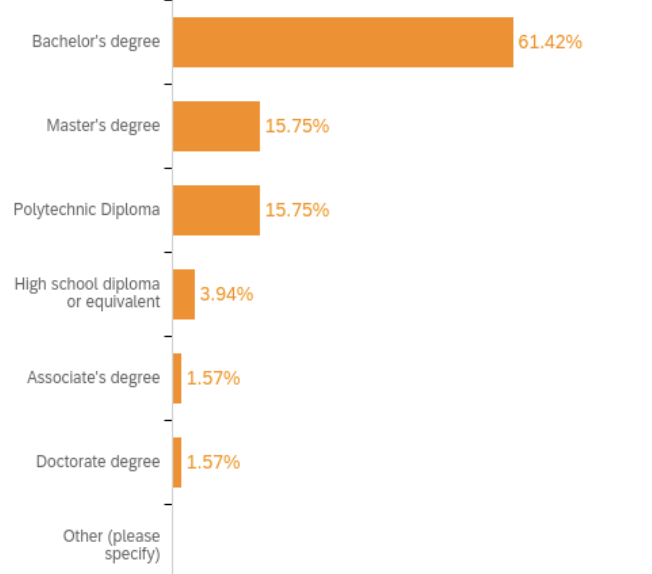}
\includegraphics[width=3in]{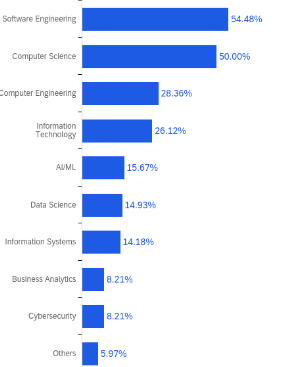}
\caption{Respondent demographics}
\label{demo}
\end{figure}
We conducted a survey among students to understand their needs, expectations, and potential adoption barriers to our DevOps learning tool. This survey helped evaluate the prototype's necessity in meeting real-world expectations. The research unfolded in three stages:

\begin{itemize}
\item \textbf{Survey Design:} The survey was meticulously crafted to obtain quantitative and qualitative data, incorporating multiple-choice, Likert scale, and open-ended questions. It covered critical areas like the tool’s ease of use, learning effectiveness, accessibility, and relevance to the current DevOps landscape, also inviting suggestions for improvement.

\item \textbf{Participant Recruitment and Survey Distribution:} A diverse group of participants were recruited, including undergraduate and postgraduate IT students and educators from various universities, to ensure findings could be generalized. Distribution channels included academic forums, email lists, and DevOps and IT education-focused social media platforms like Linkedin.

\item \textbf{Data Analysis:} Employing statistical methods for quantitative data and thematic analysis for qualitative responses, we identified patterns, trends, and insights regarding user needs, tool usability, and alignment with real-world expectations.
\end{itemize}

\begin{figure}
\centering
\includegraphics[width=3in]{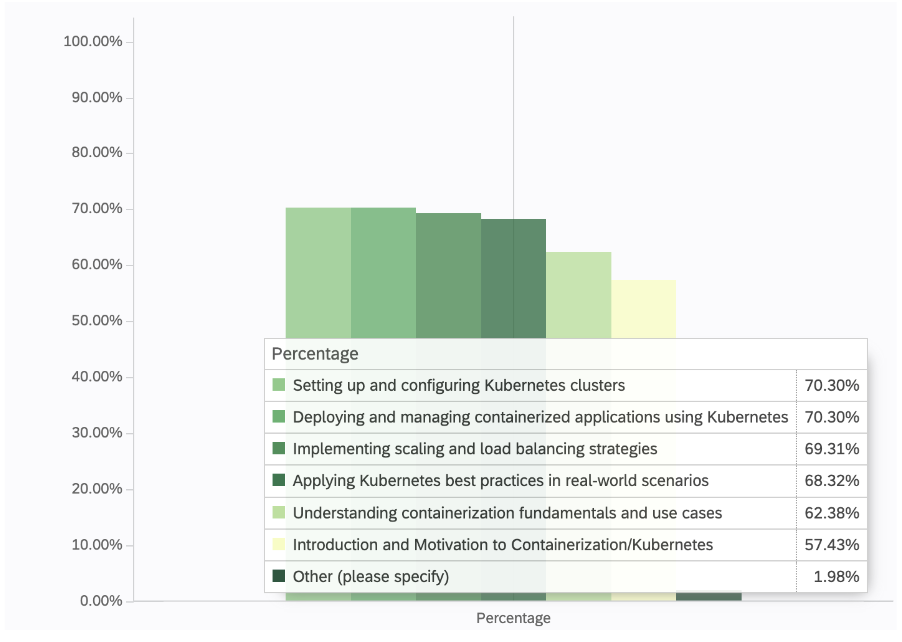}
\caption{Interest to learn Kubernetes}

\label{K8s}
\end{figure}

\begin{figure}
\centering
\includegraphics[width=3in]{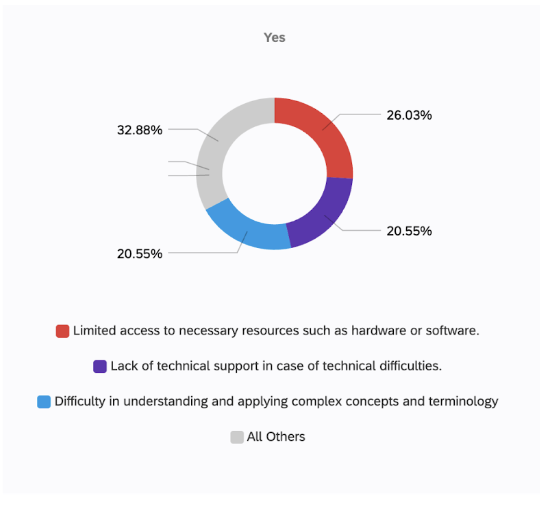}
\includegraphics[width=3in]{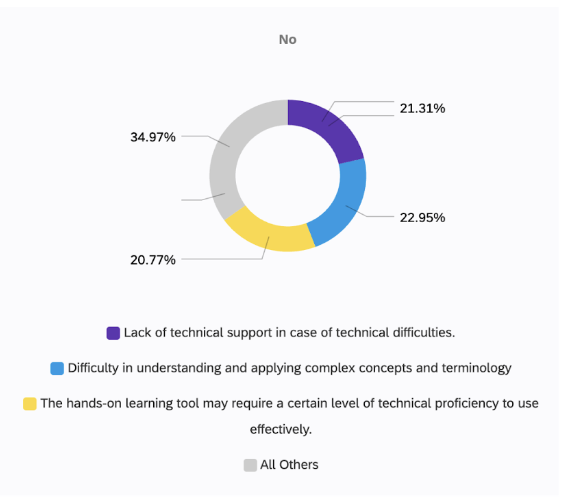}
\caption{Access to required infrastructure}
\label{infra}
\end{figure}

\begin{figure}
\centering
\includegraphics[width=3in]{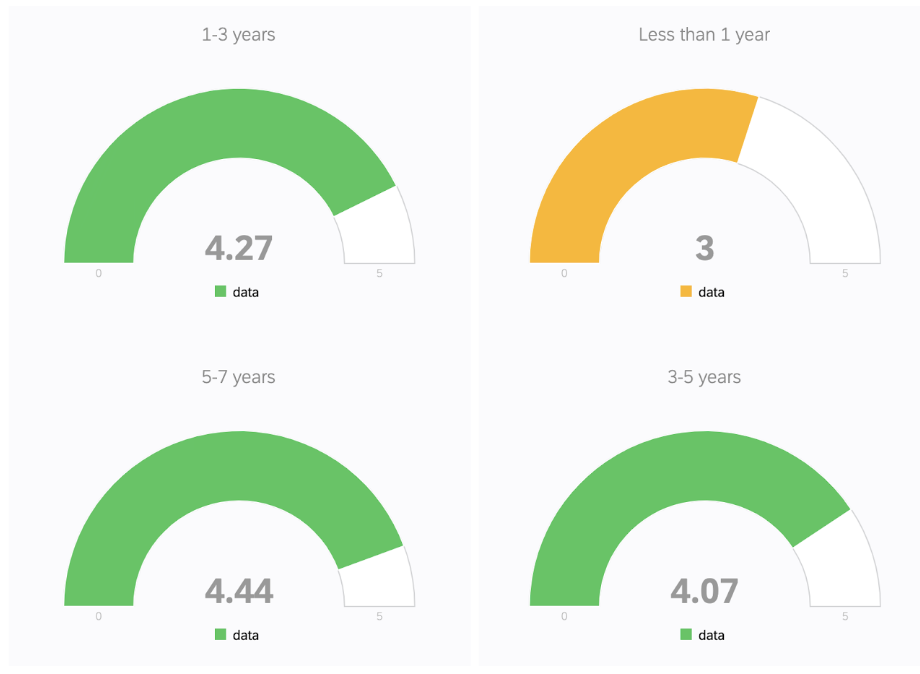}
\caption{Prior Exerience and Hands-on correlation}
\label{corr}
\end{figure}

We had 152 respondents aged between18 and 44. The feedback from the survey informed subsequent usability testing and refinements to the tool, helping to create a functional, user-friendly, and relevant solution in practical settings.
As illustrated in Fig. \ref{demo}, respondents are from diverse backgrounds - United States, Vietnam and Ukraine. 90\% of the respondents are from India and Singapore, and 60\% of respondents have a Bachelor's degree. Most respondents have 3 to 5 years of experience in computer science fields. \\ \\
\begin{figure}
\centering
\includegraphics[width=3in]{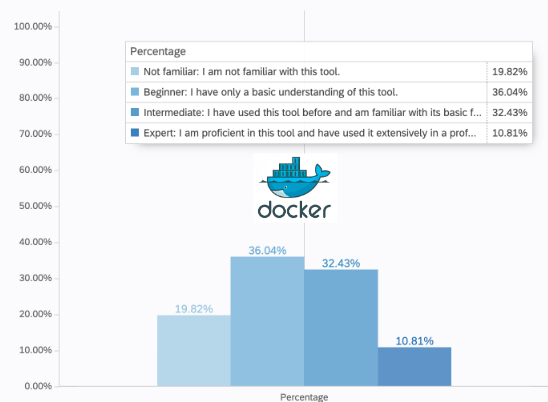}
\includegraphics[width=3in]{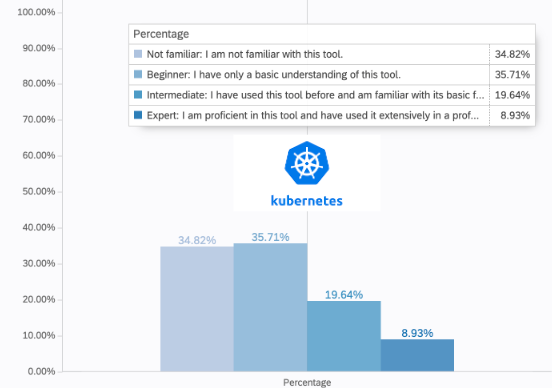}
\includegraphics[width=3in]{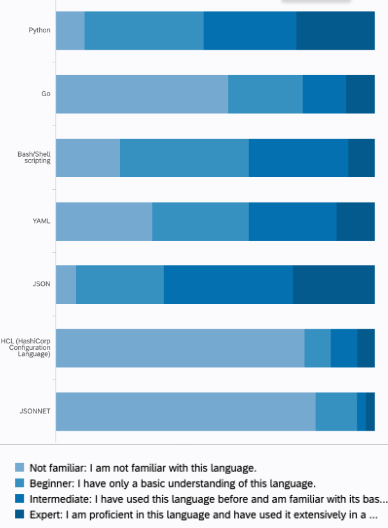}
\caption{Respondent knowledge}
\label{know}
\end{figure}
Fig. \ref{know} illustrates the participant's knowledge on some of the improtant DevOps tools. Approximately 70\% said they are beginner or intermediate users of Docker. Approximately 70\% said they are not familiar or are beginner level on Kubernetes knowledge. More students are unfamiliar with k8s than docker. When it comes to Site Reliability Engineering (SRE) tools such as JSON, YAML and Go, participants know JSON well. Students know what JSON is and consider themselves familiar with the language - with JSONNET, it builds an abstraction on top of JSON. It can also teach students YAML as the tool can convert between JSON and YAML. Their familiarity with YAML, HCL and JSONNET are lower. These are important skills needed to work with Kubernetes. \\ \\ 
As illustrated in Fig. \ref{K8s}, our survey indicated a strong student interest in learning to set up Kubernetes clusters and deploy applications. Although preferences among various topics were relatively close, this interest underscored a desire to master Kubernetes container orchestration. This insight led us to prioritize developing a focused learning module on Kubernetes cluster setup and application deployment. \\ \\
Our survey found that (refer to Fig \ref{infra}) students, especially those still studying, primarily struggled with limited access to hardware and software for learning. This highlighted the need for a cloud-based approach to overcome resource constraints. Additionally, students noted a lack of technical support, affirming our focus on creating an automated installation process. This approach is designed to ease the manual setup challenges and offer the needed technical support, facilitating a smoother learning experience. \\ \\
Our survey findings (refer to Fig \ref{corr}) show students with more experience prefer hands-on learning tools. Consequently, we aimed our project at recent graduates and professionals with some industry experience. Tailoring the tool to these users allows us to address their needs and preferences for practical learning experiences effectively, offering content that is pertinent and tailored to individuals at different career stages.

\begin{figure*}[h]
    \centering
    \includegraphics[width=1.0\linewidth]{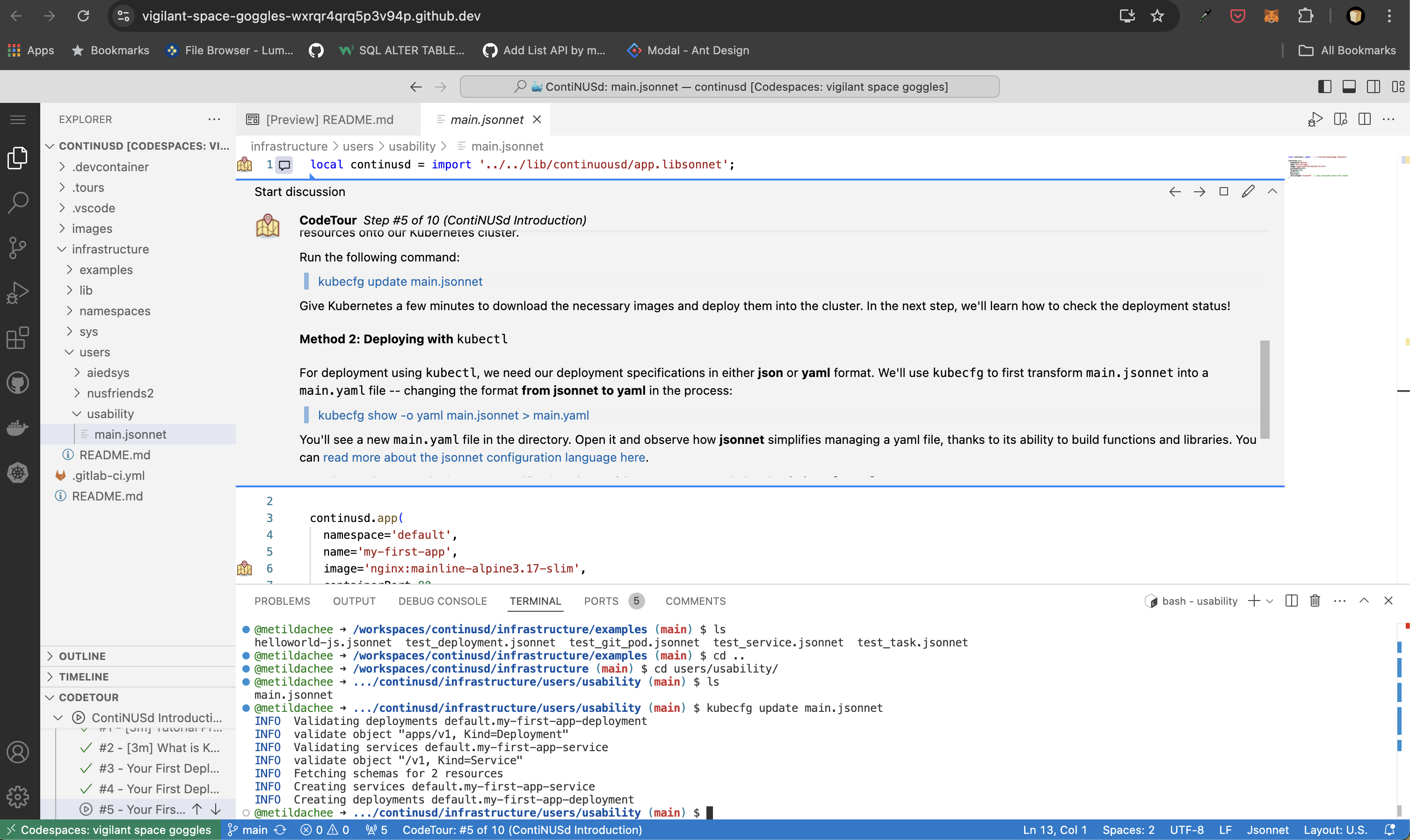}
    \caption{Standard Learning Environment}
    \label{fig:3 Standardised Learning Environment}
\end{figure*}

\subsection{Prototype Development}

Informed by our literature review and survey insights, we developed a prototype aimed at simplifying DevOps tools installation and facilitating extension. The architecture of our prototype is illustrated in Fig. \ref{Architecture Diagram}. The process included:

\begin{itemize}
    \item \textbf {Design:} Guided by the goal of simplicity and extensibility, we designed an intuitive and user-friendly interface to accommodate users with diverse skill levels.
    
    \item \textbf {Development:} The open-source prototype is relevant to the current DevOps landscape, allowing for future enhancements and adaptations.
    
    \item \textbf {Initial Testing:} Preliminary tests ensured basic functionality and allowed for early identification and resolution of issues.

    \item \textbf {Evaluation:} We assessed ease of installation, user-friendliness, extensibility, and alignment with DevOps practices to inform refinements.
\end{itemize}

\subsubsection{Creation of a Standardized Learning Environment}

We designed a standardized, reproducible online learning environment considering the diversity among students. This approach requires users only to have a browser, utilizing containerized applications and an online version of Visual Studio Code (VS Code) hosted by Github/Microsoft. This setup, which includes minikube for local Kubernetes operation and allows integration with larger cloud Kubernetes clusters as needed, provides a consistent and accessible learning environment with reduced setup complexities, meeting various learning needs and project requirements. \\ \\
The image illustrated in Fig \ref{fig:3 Standardised Learning Environment} shows Visual Studio Code (VS Code) IDE operating in a browser, illustrating our tool’s strength in facilitating seamless setup of consistent environments for students, tutors, and classmates, supporting a cohesive learning experience. Through utilizing devcontainers, our tool bundles necessary tools and software in a containerized environment, ready for immediate use. This setup demonstrates the tool's efficiency in providing hands-on learning while addressing common setup and configuration challenges.



\subsection{Prototype Evaluation}

In this project, success is defined by the tool's effectiveness in enhancing students' understanding of DevOps through hands-on experience with contemporary tools (e.g., GitLab, Kubernetes, Tekton, ELK stack, Grafana, etc.), as evidenced by improved quiz scores post-use. The tool is designed to be beginner-friendly, maintainable, and extensible, and includes mechanisms for evaluating and grading task performance, ensuring its longevity and relevance in the fast-evolving field of DevOps.

To gather user feedback, we conducted pilot studies with users.

Preceding and subsequent to tool utilisation, we administered comprehensive user surveys and anecdotal feedback. This multifaceted approach encompasses qualitative and quantitative assessments to provide better substantiated results.

The resultant data indicates a prevailing positive sentiment amongst a wide range of users, collectively indicating the efficacy and utility of the tool.

The user guide \cite{UserGuide}, demo link \cite{DemoVid} and live tool \cite{ToolLink} can be found in the reference.

\section{Architecture}
\begin{table*}[!h]
\begin{center}
\caption{\label{key-technologies-and-tools}The key technologies and tools utilised to develop the learning platform.}
\begin{tabular}{ |p{2cm}|p{10cm}| } 
 \hline
 \textbf{Tool} & \textbf{Details} \\ 
 \hline
     \par 
     Github & 
     To manage Merge Requests and facilitate code maintenance. \\ 
 \hline
  \par Kubernetes & 
    To learn Kubernetes-related skills. \\ 
 \hline
   \par JSONnet & 
   To develop libraries that conceal infrastructure complexity, enabling students to focus.
   \\ 
 \hline
    \par Dev Containers & 
    To create reproducible development environment. \\ 
 \hline
     \par Github Codespaces & Codespaces allow single-click setup for students. \\ 
 \hline
    \par CodeTour & 
    To break down concepts into structured learning. \\
\hline
\end{tabular}
\label{tab:arch}
\end{center}
\end{table*}

Our tool is an open-source client accessible through a browser via Github codespace, making it compatible with any device supporting a browser, including mobile phones, tablets, and laptops. For an authentic learning experience, the platform utilizes kubectl, a standard client for Kubernetes clusters interaction, instead of custom command-line tools. This approach provides students hands-on experience with industry-standard tools.

The system architecture is outlined below:

\paragraph{Local Environment (Student)}
The student's local environment merely needs a web browser to access ContiNUSd, simplifying the initial setup process.

\paragraph{Remote Server (Codespaces, VS Code Web, VS Code Server)}
Hosted remotely, this section consists of:

\paragraph{Codespaces}
Codespaces offers a cloud-based environment facilitating the easy launch and management of learning instances while handling the necessary setup and configurations.

\paragraph{VS Code Web}
This browser version of Visual Studio Code (VS Code) provides a familiar interface for students to engage in coding activities seamlessly.

\paragraph{VS Code Server}
Serving as the backend for VS Code Web, the server manages VS Code Web instances and delivers the IDE's core functionalities.

\paragraph{Devcontainer}
Operating within the VS Code Server, Devcontainer creates a consistent development environment. It can run a local Kubernetes environment or connect to a remote cluster, offering a flexible learning environment equipped with the necessary tools and configurations for smooth Kubernetes interaction.

\begin{figure}[h]
\centering
\includegraphics[scale=0.5]{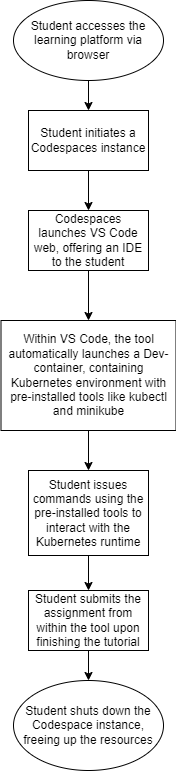}
\caption{Sample workflow}
\label{fig:6 Example Workflow}
\end{figure}

This architecture summarized in Table \ref{tab:arch} allows students to effortlessly access the tool via a browser, with the remote server managing the backend components. The arrangement ensures a standardized, configurable, and consistent learning experience irrespective of the student's device while providing essential resources and tools for effective Kubernetes learning.
A sample workflow of a student interacting with the tool is illustrated in Fig. \ref{fig:6 Example Workflow}.

\section{Conclusion}
This paper highlights the importance of DevOps in software engineering education, introducing a tested and validated tool that facilitates DevOps learning for students and educators. The presented online IDE simplifies the setup process while adapting to industry changes, providing a standardized, extendable, browser-based learning environment with key tools like VS Code and Kubernetes, ensuring students gain practical, industry-aligned experience.

Feedback from students and DevOps professionals confirms the tool’s effectiveness in bridging academic knowledge and industry expectations. This response underscores the need for educational approaches that proactively respond to the evolving software industry. In summary, our study advocates for and demonstrates the value of innovative, practical learning tools in the dynamic field of DevOps education.



\bibliographystyle{IEEEtran}
\bibliography{continusd}

\end{document}